\documentclass[prd,twocolumn,showpacs,amsmath,amssymb]{revtex4-1}
\usepackage{graphicx}
\usepackage{epsfig,graphics,subfigure,psfrag,amsmath,amssymb}
\usepackage{lineno}
\usepackage{dcolumn}
\usepackage{ulem}
\usepackage{bm}
\usepackage{overpic}
\usepackage{tabularx}
\usepackage{color}
\usepackage{multirow}
\usepackage{colortbl}
\usepackage{epstopdf}
\usepackage{appendix}
\usepackage{rotating}
\usepackage{sidecap}
\usepackage{dcolumn}
\usepackage[colorlinks,linkcolor=blue,anchorcolor=blue,citecolor=blue,urlcolor=blue]{hyperref}
\def\BR{\mathcal{B}}
\def\Etaf{\eta_{\gamma\gamma}}
\def\Etas{\eta_{\pi^0\pi^+\pi^-}}
\def\Etapf{\eta^{\prime}_{\eta_{\gamma\gamma}\pi^+\pi^-}}
\def\Etaps{\eta^{\prime}_{\gamma\rho^0}}
\def\bkgEta{D_s^+\rightarrow \eta e^+\nu_e}

\def\sigGamma{D_s^+\rightarrow \gamma e^+\nu_e}
\def\result{ 1.3\times10^{-4}}
\def\resultb{5.7\times10^{-5}}

\begin{document}
\title{Search for the decay  $D_s^+ \rightarrow \gamma  e^{+} \nu_e$}

\author{
\small
M.~Ablikim$^{1}$, M.~N.~Achasov$^{10,d}$, S. ~Ahmed$^{15}$, M.~Albrecht$^{4}$, M.~Alekseev$^{55A,55C}$, A.~Amoroso$^{55A,55C}$, F.~F.~An$^{1}$, Q.~An$^{52,42}$, Y.~Bai$^{41}$, O.~Bakina$^{27}$
% ...
\\
\vspace{0.2cm}
(BESIII Collaboration)\\
\vspace{0.2cm} {\it
$^{1}$ Institute of High Energy Physics, Beijing 100049, People's Republic of China\\
% ...
$^{4}$ Bochum Ruhr-University, D-44780 Bochum, Germany\\
% ...
$^{10}$ G.I. Budker Institute of Nuclear Physics SB RAS (BINP), Novosibirsk 630090, Russia\\
% ...
$^{15}$ Helmholtz Institute Mainz, Johann-Joachim-Becher-Weg 45, D-55099 Mainz, Germany\\
% ...
$^{27}$ Joint Institute for Nuclear Research, 141980 Dubna, Moscow region, Russia\\
% ...
$^{41}$ Southeast University, Nanjing 211100, People's Republic of China\\
$^{42}$ State Key Laboratory of Particle Detection and Electronics, Beijing 100049, Hefei 230026, People's Republic of China\\
% ...
$^{52}$ University of Science and Technology of China, Hefei 230026, People's Republic of China\\
% ...
$^{55}$ (A)University of Turin, I-10125, Turin, Italy; (B)University of Eastern Piedmont, I-15121, Alessandria, Italy; (C)INFN, I-10125, Turin, Italy\\
% ...
$^{d}$ Also at the Novosibirsk State University, Novosibirsk, 630090, Russia\\
% ...
}
}



\begin{abstract}
A search for the rare radiative leptonic decay $D_s^+\to\gamma e^+\nu_e$ is performed for the first time using electron-positron collision data corresponding to an integrated luminosity of 3.19 fb$^{-1}$, collected with the BESIII detector at a center-of-mass energy of 4.178 GeV. No evidence for the $D_s^+\to\gamma e^+\nu_e$ decay is seen and an upper limit of $\mathcal B(D_s^+\to\gamma e^+\nu_e)<1.3\times 10^{-4}$ is set on the partial branching fraction at a 90\% confidence level for radiative photon energies $E_{\gamma}^*>0.01$~GeV.
\end{abstract}

\maketitle


\section{Introduction}

\begin{figure}
  \mbox{
    \begin{overpic}[width=4cm,angle=0]{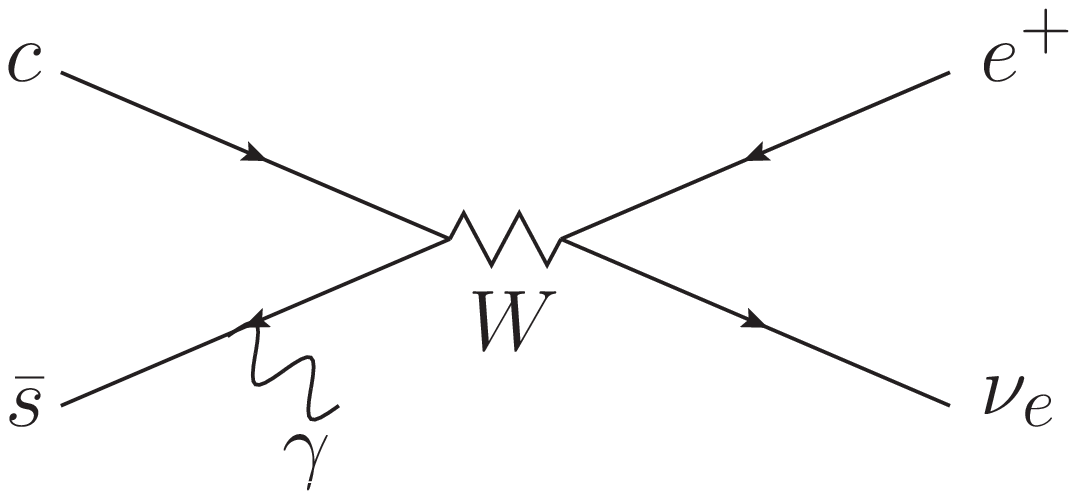}
      \put(42,-0){{\small }}
    \end{overpic}
    \begin{overpic}[width=4cm,angle=0]{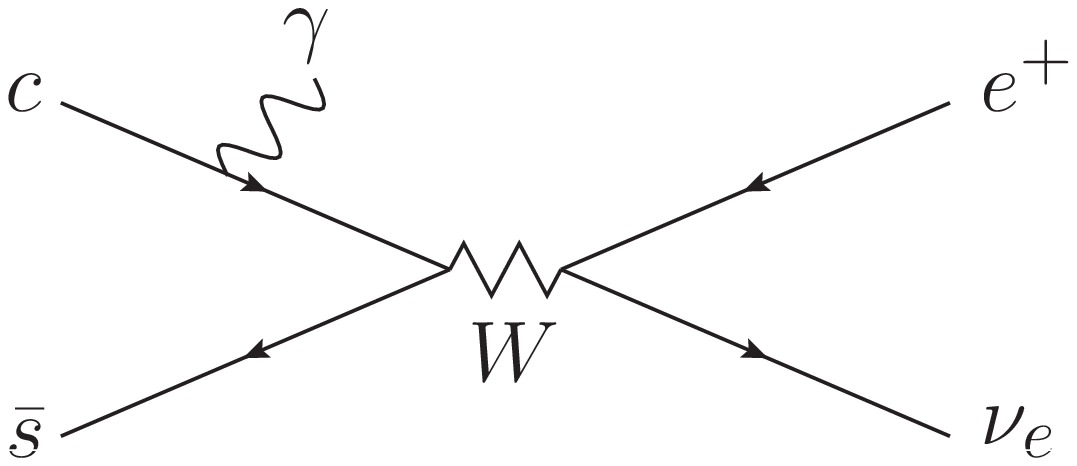}
      \put(42,-0){{\small  }}
    \end{overpic}
  }
  \mbox{
    \begin{overpic}[width=4cm,angle=0]{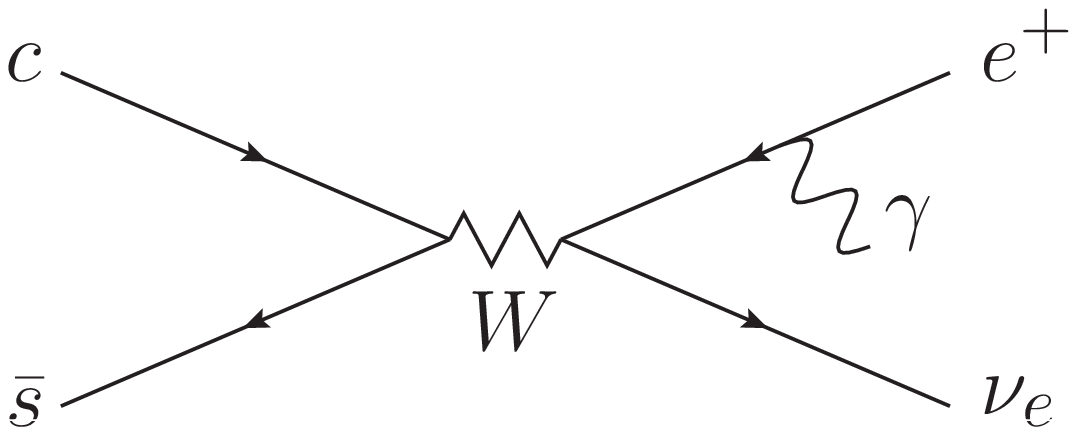}
      \put(42,-0){{\small  }}
    \end{overpic}
    \begin{overpic}[width=4cm,angle=0]{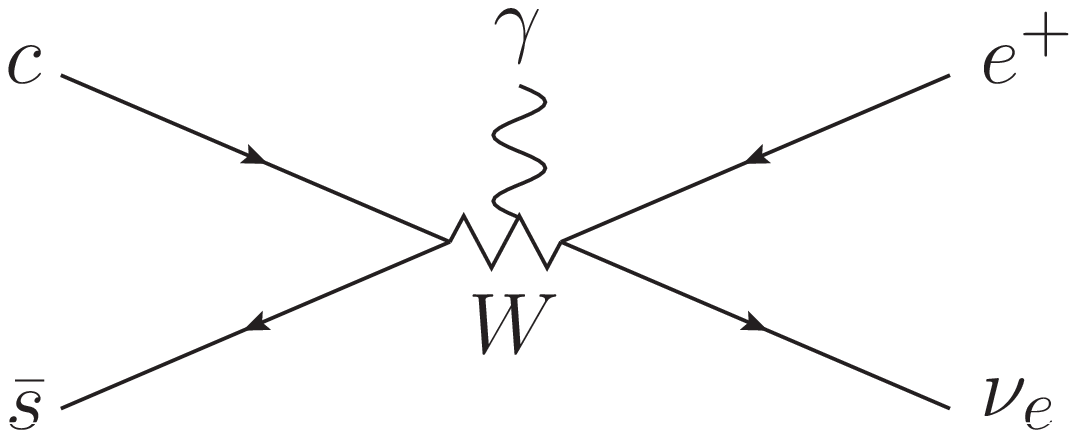}
      \put(42,-0){{\small  }}
    \end{overpic}
  }
  \caption{Tree-level Feynman diagrams contributing to $D_s^+\to\gamma e^+\nu_e$.\label{fig:tree}}
\end{figure}

In the Standard Model, the purely leptonic decays of heavy pseudoscalar mesons, $P\to e^+\nu_e$, are helicity suppressed by a factor $m_e^2$.
The helicity suppression in these processes can be overcome by the emission of a radiative photon as shown in Fig.~\ref{fig:tree}. As a result, the decay rate of the purely leptonic radiative decay $P\to \gamma e^+ \nu_e$ may be $10^3-10^5$ times~\cite{Richman:1995wm} larger than that of $P\to e^+ \nu_e$.
For example, the branching fractions (BFs) of $D^+_{(s)}\to\gamma e^+\nu_e$ are theoretically predicted to range from $10^{-5}$ to $10^{-3}$~\cite{Geng:2000if,Korchemsky:1999qb,Lu:2002mn,Yang:2012jp,Yang:2014rna,Atwood:1994za,Burdman:1994ip}.
An experimental search for these decays can shed light on the dynamics of the underlying processes and can provide input of decay rates to theoretical calculations.

\begin{figure}
  \centering
  \includegraphics[width=8cm]{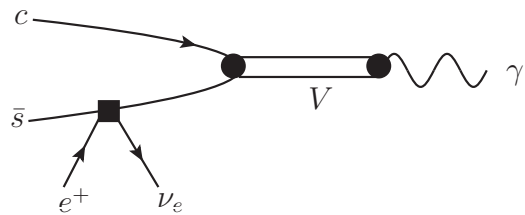}
  \caption{Long-distance contribution to the radiative leptonic decays proceeds via a semileptonic intermediate state, $e^+\nu_eV$, where $V$ can be a $\rho$, $\omega$ or a $\phi$ meson, and $V$ turns into an on-shell photon $V\to\gamma$~\cite{Yang:2012jp}.\label{fig:feynman}}
\end{figure}

Previously, the BESIII experiment has searched for the radiative leptonic decay $D^+\to\gamma e^+\nu_e$ using a data sample collected at a center-of-mass energy  $\sqrt{s}=3.773$~GeV. No significant signal is observed and an upper limit on the partial decay BF for radiative photon energies $E_\gamma^*>0.01$~GeV is set to $\mathcal{B}< 3.0 \times 10^{-5}$ at the 90\% confidence level (C.L.)~\cite{Ablikim:2017twd}, approaching the range of  theoretical predictions, $(1.9$-$2.8)\times10^{-5}$~\cite{Yang:2012jp,Yang:2014rna}.
The decay $D^+\to\gamma e^+\nu_e$ is Cabibbo suppressed, while the decay $D_s^+\to\gamma e^+\nu_e$ is Cabibbo favored. The full BF of  $D_s^+\to\gamma e^+\nu_e$ is predicted to be of the order $10^{-5}$-$10^{-4}$ in the light front quark model~\cite{Geng:2000if} and in the nonrelativistic constituent quark model~\cite{Lu:2002mn}.  The theoretical study in Ref.~\cite{Yang:2012jp} indicates that the long-distance contribution described by the vector meson dominance model, as shown in Fig.~\ref{fig:feynman}, may further enhance this decay BF up to order $10^{-4}$.  Moreover the BF is predicted to be of order $10^{-3}$ within the perturbative
quantum chromodynamics method combining heavy quark effective theory~\cite{Korchemsky:1999qb}. With a BF sensitivity of $10^{-4}-10^{-5}$, this decay may be detectable at BESIII.

In this paper, we report on the first search for the radiative leptonic decay $D_s^+\to\gamma e^+\nu_e$, using a data sample corresponding to an integrated luminosity of 3.19~fb$^{-1}$ of $e^+e^-$ collisions collected at $\sqrt{s} = 4.178$~GeV with the BESIII detector in 2016. To reduce the risk of bias, the analysis procedure of the nominal analysis has been developed as a blind analysis, based on an inclusive Monte Carlo~(MC)-simulated data sample with equivalent luminosity the same as data. The inclusion of the charge conjugate process is implied throughout the paper unless explicitly specified otherwise.

\section{ BESIII DETECTOR AND DATA SET}

The BESIII detector is a magnetic
spectrometer~\cite{Ablikim:2009aa} located at the Beijing Electron
Positron Collider (BEPCII)~\cite{Yu:IPAC2016-TUYA01}. The
cylindrical core of the BESIII detector consists of a helium-based
 multilayer drift chamber (MDC), a plastic scintillator time-of-flight
system (TOF), and a CsI (Tl) electromagnetic calorimeter (EMC),
which are all enclosed in a superconducting solenoidal magnet
providing a 1.0~T magnetic field. The solenoid is supported by an
octagonal flux-return yoke with resistive plate counter muon
identifier modules interleaved with steel. The acceptance of
charged particles and photons is 93\% over $4\pi$ solid angle. The
charged particle momentum resolution at $1~{\rm GeV}/c$ is
$0.5\%$, and the specific energy loss ($dE/dx$) resolution is $6\%$ for the electrons
from Bhabha scattering. The EMC measures photon energies with a
resolution of $2.5\%$ ($5\%$) at $1$~GeV in the barrel (end cap)
region. The time resolution of the TOF barrel part is 68~ps. The end cap TOF
system was upgraded in 2015 with multi-gap resistive plate chamber
technology, providing a time resolution of
60~ps~\cite{Lxin,Gyingxiao}.

MC-simulated events are generated with the {\sc{geant4}}-based~\cite{Agostinelli:2002hh} software package {\sc boost}~\cite{Deng} that describes the detector geometry and material, implements the detector response, simulates digitization, and incorporates time-dependent beam backgrounds. An inclusive simulation sample, which includes open charm processes, the initial-state radiation (ISR) production of $\psi(3770)$, $\psi(3686)$ and $J/\psi$,
  $q\bar q (q=u,\,d,\,s$) continuum processes, along with Bhabha scattering, $\mu^+\mu^-$, $\tau^+\tau^-$ and $\gamma\gamma$ processes, is produced at $\sqrt{s} = 4.178$~GeV. The open charm processes are simulated using
  {\sc{conexc}}~\cite{Ping:2013jka}. The effects of~ISR and final-state radiation (FSR)~\cite{RichterWas:1992qb} are taken into account. Decays of unstable particles are simulated by {\sc{evtgen}}~\cite{Ping:2008zz} using  BFs  from the Particle Data Group~\cite{Patrignani:2016xqp},
and the remaining unknown decay  modes of $\psi$ are generated using the modified {\sc{lund}} model~\cite{Chen:2000tv}.
The signal candidates are simulated using the method employed in Ref.~\cite{Ablikim:2017twd}, where the two parameters, the decay constant~\cite{Patrignani:2016xqp} and the quark mixing matrix element~\cite{Patrignani:2016xqp}  are adjusted according to the decay channel. The minimum energy of the radiative photon of the $D^+_s\to \gamma e^+\nu_e$ decay is set at 0.01~GeV to avoid the infrared divergence for soft photons.

\section{ DATA ANALYSIS}

At $\sqrt{s} = 4.178$~GeV, the $D_s$ mesons are mostly produced in the process
$e^+e^-\to D^+_sD_s^{*-}$.  This allows us to perform the analysis using a modified double-tag~(DT) technique~\cite{Baltrusaitis:1985iw}. First, the $D_s^-$ decay is fully reconstructed, leading to the single-tag~(ST) mesons. The ST candidates that contain the signal decay $D_s^+\to\gamma e^+\nu_e$, which are called the DT events are  selected and investigated in the presence of one additional isolated photon or $\pi^0$ meson originating from the $D_s^{*}$ decay. The BF of~$D_s^+\to\gamma e^+\nu_e$ is determined by
\begin{equation}
\BR(D_s^+\to\gamma e^+\nu_e) = \frac{N_{\rm signal}}{N^{\rm tot}_{\rm ST}\epsilon_{\gamma_{\rm soft}(\pi^0_{\rm soft})\rm SL}},
\end{equation}
where $N^{\rm tot}_{\rm ST}$ and $N_{\rm signal}$ are the ST and DT yields in data, respectively. $\epsilon_{\gamma_{\rm soft}(\pi^0_{\rm soft})\rm SL}$ is the reconstruction efficiency for ``$\gamma_{\rm soft}(\pi^0_{\rm soft})D_s^+,~D_s^+\to\gamma e^+\nu_e$" determined by
$\sum_i\frac{N^i_{\rm ST}}{N^{\rm tot}_{\rm ST}}\frac{\epsilon^i_{\rm DT}}{\epsilon^i_{\rm ST}} $,
where $\gamma_{\rm soft}(\pi^0_{\rm soft})$ denotes the soft $\gamma$ or $\pi^0$ from the $D^{*-}_s$, $\gamma e^+\nu_e$ decays come from either the bachelor $D_s^+$ or $D_s^{*+}$,  $\epsilon^i_{\rm ST}$ and $\epsilon^i_{\rm DT}$ are the efficiencies of selecting the ST and DT candidates, and $i$ denotes the $i$ th tag mode as described below.

The ST candidates are reconstructed through the decay modes $D_s^-\to K^+K^-\pi^-$,~$K^+K^-\pi^-\pi^0$,~$K^0_SK^-$, $\Etaf\pi^-$, $\Etas\pi^-$, $\pi^+\pi^-\pi^-$, $K_S^0K^+\pi^-\pi^-$, $K_S^0K^-\pi^+\pi^-$, $\Etapf\pi^-$, $\Etaps\pi^-$, $K_S^0K_S^0\pi^-$, $K_S^0K^-\pi^0$,
$K^-\pi^+\pi^-$ and $\Etaf\rho^-$, where the subscripts of $\eta^{(\prime)}$ represent the decay modes used to reconstruct $\eta^{(\prime)}$. All charged tracks must have a polar angle ($\theta$) within
$|\cos\theta|< 0.93$. The reconstructed tracks are required to point back to the interaction point~(IP) region with $|V_{r}|<1$~cm and $|V_{z}|<10$~cm, where $|V_{r}|$ and $|V_{z}|$ are the distances of closest approach to the IP
in the transverse plane and along the positron beam direction, respectively.  Charged kaons and pions are identified by using the combined information from $dE/dx$ and TOF.
The charged tracks are assigned as pion~(kaon) candidates if ${\mathcal L}_{\pi(K)}>{\mathcal L}_{K(\pi)}$, where $\mathcal{L}_{\pi(K)}$ is the C.L. for the pion (kaon) hypothesis.  Below 1.2 GeV/$c$, the particle identification~(PID) efficiencies of charged kaons (pions) range from 89\%\,(85\%) to 99\%, while the rates of misidentifying kaons (pions) as pions (kaons) range from 1\% to 12\%\,(15\%).

The $K_S^0$ candidates
are formed from pairs of oppositely charged tracks satisfying $|V_{z}|<20$~cm. The two charged tracks are taken as $\pi^+\pi^-$ without identification requirements and are constrained to have a common vertex. The invariant mass of the $\pi^+\pi^-$ pair is required to be within (0.487,\,0.511)~GeV/$c^2$. The decay length of the $K^0_S$ candidate is required to be larger than twice the vertex resolution away from the IP.

Photon candidates are reconstructed from clusters of energy deposited in
the EMC, with the energy measured in nearby TOF counters included to improve reconstruction efficiency and energy resolution.~
The energies of photon candidates must be larger than 0.025~(0.05)~GeV for the barrel~(end cap) region. These requirements are safe for the minimum energy requirement $E_\gamma^* > 0.01$~GeV on the radiative photon.
The cluster timing~\cite{StartTime} is required to be between 0 and 700~ns to suppress electronic noise and energy depositions unrelated to the event of interest.

Pairs of photon candidates are combined to form $\pi^0\to\gamma\gamma$ and $\eta\to\gamma\gamma$ candidates, and a kinematic fit constraining the $\gamma\gamma$ invariant mass to the corresponding nominal mass is performed to improve the four-momentum resolution. The $\pi^0$ and $\eta$ candidates are selected with their unconstrained $\gamma\gamma$ masses within (0.115,~0.150) and (0.50,~0.57) GeV/$c^2$, respectively. We reconstruct $\eta\to\pi^+\pi^-\pi^0$ candidates by requiring $M_{\pi^0\pi^+\pi^-}\in(0.53,\,0.57)$~GeV/$c^2$.

We select $\eta^\prime$ candidates in two final states: $\eta_{\gamma\gamma}\pi^+\pi^-$ and $\gamma\pi^+\pi^-$. The invariant mass of the reconstructed $\eta^\prime$ candidate is required to satisfy $M_{\eta_{\gamma\gamma}\pi^+\pi^-}\in(0.946,\,0.970)$~GeV/$c^2$ or $M_{\gamma\rho^0}\in(0.940,\,0.976)$~GeV/$c^2$.

To remove the soft pions coming from  $D^*$ decay, the momentum of the pion coming directly from the ST $D_s^-$ decay must be larger than 0.1 GeV/$c$.
For the $\pi^+\pi^-\pi^-$ and $K^-\pi^+\pi^-$ final states, the contributions of $D^-_s\to K^0_S\pi^-$ and $K^0_SK^-$ are rejected if $M_{\pi^+\pi^-}$ lies within $\pm$0.03~GeV/$c^2$ of the nominal $K^0_S$ mass~\cite{Patrignani:2016xqp}.
 \begin{figure*}[htp]
  \centering
  \includegraphics[width=16cm]{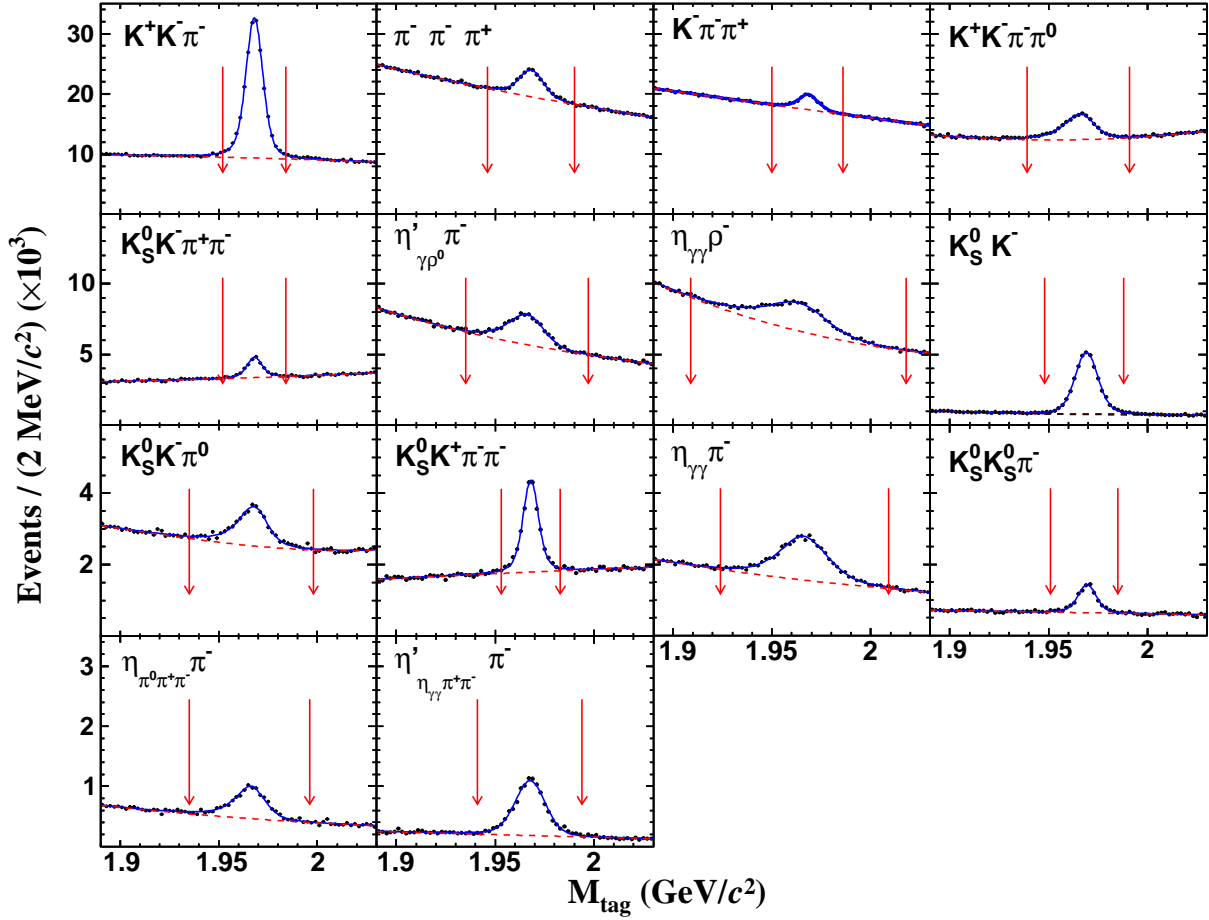}
  \caption{\label{fig:ST}\small Reconstructed mass $M_\text{tag}$ of the selected single tag events. Superimposed on the data points in black is the signal and background combined fit (solid blue line); the dashed red line describes the combinatorial background, and the dashed black line in the $K_S^0K^-$ mode corresponds to the $D^-\to K_S^0 \pi^-$ background contribution. The arrows indicate the definition of the $D_s^-$ signal region.}
  \end{figure*}

  The ST $D^-_s$ mesons are identified by the modified mass
  \begin{equation}
  M_{\rm  mod}\equiv\sqrt{E^2_{\rm beam}- |\vec{p}_{D_s^-}|^2}
  \end{equation} and the $D_s^-$ recoil mass
  \begin{eqnarray}
 \nonumber M_{\rm rec} \equiv
 \sqrt{\left({2E_{\rm beam} - \sqrt{|\vec{p}_{D_s^-}|^2 + M_{D_s^-}^2}}\right)^2- |\vec{p}_{D_s^-}|^2} ,
  \end{eqnarray}
  where $\vec{p}_{D^-_s}$ is the three-momentum of the ST candidate in the rest frame of the $e^+e^-$ system, $M_{D^-_s}$ is the nominal $D^-_s$ meson mass~\cite{Patrignani:2016xqp} and $E_{\rm beam}$ is the beam energy.
 The non-$D^+_s D^{*-}_s$ events are suppressed by requiring $M_{\rm mod}\in(2.010,\,2.073)$~GeV/$c^2$. In each event, only the candidate with the $M_{\rm rec}$ closest to the $D_s^{*+}$  nominal mass~\cite{Patrignani:2016xqp} is chosen. The  invariant mass~($M_{\rm tag}$) spectra
  of the accepted ST candidates for the 14 tag modes are shown in Fig.~\ref{fig:ST}. The ST yield is determined via unbinned maximum-likelihood
fits to each spectrum. Signals and the $D^-\to K^0_S\pi^-$ peaking background with a tiny fraction (dashed black line in Fig.~\ref{fig:ST}) in the $D^-_s\to K^0_SK^-$ mode are described by MC-simulated shapes using the kernel density estimation method~\cite{Cranmer:2000du}. To take into account the resolution difference between data and simulation, the MC-simulated shapes are convolved with a Gaussian function for each tag mode, where the parameters of the Gaussian function are left free in the fit. The nonpeaking background is modeled by a second- or third-order Chebychev polynomial function, and the reliability of the fitted nonpeaking background has been verified using the inclusive MC sample.
     Candidates in the signal regions, denoted by the boundaries in each subfigure of Fig.~\ref{fig:ST}, are kept for further analysis. The $M_{\rm tag}$ signal regions, the ST yields in data and the ST efficiencies are summarized in Table~\ref{tab:ST}. The total ST yield is $N^{\rm tot}_{\rm ST}=395412\pm1931$, where the uncertainty is statistical.

\begin{table*}
            \caption{Summary of the $M_{\rm tag}$ mass windows, ST yields of data~($N_{\rm ST}$), ST~($\epsilon_{\rm ST}$) and DT~($\epsilon_{\rm DT}$) efficiencies. All uncertainties are statistical only.\label{tab:ST}}
              \centering
                \begin{tabular}{lcr@{$\pm$}cc@{$\pm$}cr@{$\pm$}c}\hline\hline
               Mode                          & $M_{\rm tag}$ (GeV/$c^2$) &\multicolumn{2}{c}{$N_{\rm ST}$} 	& \multicolumn{2}{c}{$\epsilon_{\rm ST}$~(\%) } & \multicolumn{2}{c}{$\epsilon_{\rm DT}$~(\%)}   \\\hline%
                 $K^+K^-\pi^-$		&	(1.952,\,1.984)	&	134679 &	561	& 	39.86  & 0.08& 17.89  & 0.06 \\
		$\pi^+\pi^-\pi^-$	&	(1.946,\,1.990)	&	  36258&	776	& 	51.73  & 0.43& 23.16 &   0.85\\
		$K^-\pi^+\pi^-$		&	(1.950,\,1.986)	&	  15540&	839	& 	44.40  & 0.58&  22.21 & 1.08 \\
		$K^+K^-\pi^-\pi^0$	&	(1.939,\,1.991)	&	  44108&	966	& 	12.28 & 0.09&    5.43 & 0.19 \\
		$K_S^0K^-\pi^+\pi^-$	&	(1.952,\,1.984)	&  7304&	243	& 	17.31  & 0.27& 5.83  & 0.36\\
		$\Etaps\pi^-$		&	(1.935,\,1.997)	&	  24602&	481	& 	29.33  & 0.26&  12.92 & 0.54\\
		$\Etaf\rho^-$		&	(1.912,\,2.016)	&	  36363&	684	& 	19.55  & 0.14&  10.53  &0.28\\
		$K^0_SK^-$		&	(1.948,\,1.988)	&	      32229&	235	&	49.85  & 0.18& 17.54 &  0.69 \\
		$K_S^0K^-\pi^0$		&	(1.935,\,1.998)	&11644&	361	& 	18.50  & 0.28&  8.91  & 0.34\\
		$K^0_SK^+\pi^-\pi^-$	&	(1.953,\,1.983)	&13780&	210	& 	19.89  & 0.15&  15.90  & 0.80\\
		$\Etaf\pi^-$		&	(1.924,\,2.009)	&	  19187&	320	& 	48.93  & 0.30&  22.42 &    0.94 \\
		$K_S^0K_S^0\pi^-$		&	(1.951,\,1.985)	&  4883&	133	& 	20.89  & 0.26& 11.32  & 0.52 \\
		$\Etas\pi^-$		&	(1.935,\,1.996)	&	    5463&	138	& 	24.31  & 0.27&  11.80 &  0.91\\
		$\Etapf\pi^-$		&	(1.941,\,1.994)	&	    9103&	131	& 	22.34  & 0.15&  10.93 & 0.66\\			
\hline\hline
 \end{tabular}
            \label{tab:tag_yields}
        \end{table*}

   The $D_s^+\to\gamma e^+\nu_e$ candidates are selected from the remaining charged tracks and showers in the side recoiling against the ST $D_s^-$ meson and the isolated photon or $\pi^0$ meson with the same criteria as used in the ST candidate selection. It is required that there be only one good charged track, with charge opposite to the ST $D_s^-$ meson. The positron is identified using the C.L. computed by combining PID information from $dE/dx$, TOF, and EMC. Under the assumption that the charged track in the signal decay is a positron, a pion, or a kaon, three C.L.s are calculated: ${\mathcal L}^\prime_e$, ${\mathcal L}^\prime_\pi$  and ${\mathcal L}^\prime_K$. The charged track is identified as a positron if $\mathcal{L}^{\prime}_e> 0.001$ and ${\mathcal L}^\prime_e/(\mathcal{L}^\prime_e + \mathcal{L}^\prime_\pi + \mathcal{L}^\prime_K) > 0.8$.   To reduce the rate of misidentifying a pion as a positron, the ratio $E_e/p_e$ is required to be greater than 0.8, where $E_e$ and $p_e$ are the deposited energy of the positron in the EMC and the momentum measured by the MDC, respectively. Below 1.2 GeV/$c$, the PID efficiencies of $e^\pm$ are greater than 98\%, while the averaged rate of misidentifying $K^\pm$ or $\pi^\pm$ as $e^\pm$ is about 0.3\%.

   To improve the degraded momentum resolution of the electron due to FSR and bremsstrahlung effects, the energies of neighboring photons are added back to the positron candidates. Specifically, the photons with energy greater than 0.03~GeV and within a cone of 5$^\circ$ around the positron direction (but excluding the radiative photon candidate) are included.

\begin{figure}[htp]
     \centering
  \includegraphics[width=8cm]{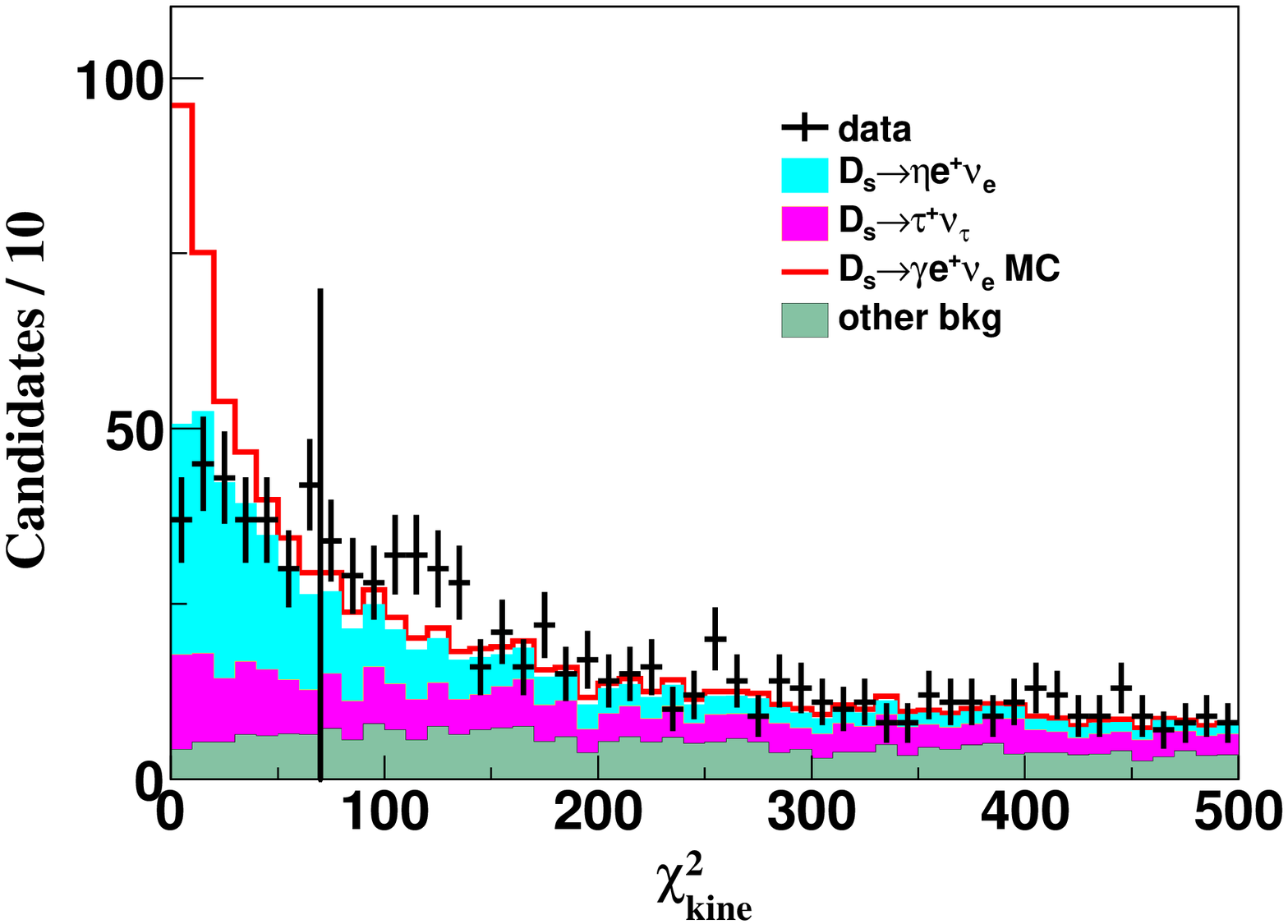}
  \caption{\label{fig:chisq} Distribution of $\chi^2_{\rm kine}$ for the selected $D_s^+\to\gamma e^+\nu_e$ candidates. The black points with error bars represent the data. The solid red curve is from the simulated signal candidates normalized with a partial BF $\BR(D_s^+\to\gamma e^+\nu_e)= 7.5\times10^{-4}$.}
          \end{figure}

  To select the radiative leptonic decay candidate from the process $e^+e^-\to D_s^{+}D_s^{*-}\to D_s^{+}D_s^{-}\gamma_{\rm soft}(\pi^0_{\rm soft})$,  we perform kinematic fits imposing four-momentum conservation under the four hypotheses of
   	 $e^+e^-\to {D^{+}_s}_{\,\gamma e^+\nu_e}{D^{*-}_s}_{\,D_s^- \gamma_{\rm soft}}$,
	${D^{+}_s}_{\,\gamma e^+\nu_e}{D^{*-}_s}_{\,D_s^- \pi^0_{\rm soft}}$,
	$D_s^{+} {D^{*-}_s}_{\gamma e^- \bar\nu_e \gamma_{\rm soft}}$, and~$D_s^{+} {D^{*-}_s}_{\,\gamma e^- \bar\nu_e \pi^0_{\rm soft}}$,~where the subscripts of $D_s^{(*)}$ represent the particle combinations of $D_s^{(*)}$.
   The ST $D_s^-$ candidates are indirectly produced from $D_s^{*-}$ in the first two hypotheses, but are directly produced from $e^+e^-$ annihilations in the latter two hypotheses. The $\gamma_{\rm soft}\,(\pi^0_{\rm soft})$ candidates from $D^{*-}$ are found in the first and third (second and fourth) hypotheses.
   The $D^{\pm}_s$ and $D^{*\pm}_s$ candidates are constrained to their individual nominal masses~\cite{Patrignani:2016xqp}. In addition, the neutrino is treated as a missing particle in the DT event.
  The hypothesis with the smallest $\chi^2_{\rm kine}$ is chosen. The $\chi^2_{\rm kine}$ distribution of the accepted candidates is shown in Fig.~\ref{fig:chisq}.

 To suppress the background from $D^+_s$ hadronic decays due to fake photons and charged tracks, the maximum energy of the showers not used in the DT event selection ($E_{\rm \gamma~extra }^{\rm max}$) is required to be less than 0.2~GeV, and events with additional charged tracks ($N^{\rm extra}_{\rm char}$) are removed. To suppress backgrounds from $D_s^+ \to \tau^+\nu_\tau$ and $D_s^+\to\eta e^+\nu_e$, $\chi^2_{\rm kine}$ is required to be less than 70. The backgrounds from $D^+_s\to\eta e^+\nu_e$ are further suppressed by rejecting the events if the invariant mass of any $\gamma\gamma$ combination that has not been used in ST selection satisfies  $M_{\gamma\gamma}\in$(0.51,\,0.56)~GeV/$c^2$. These requirements keep 80\% of the signal events, but remove more than 70\% of the background events.

\begin{figure}
     \centering
  \includegraphics[width=8cm]{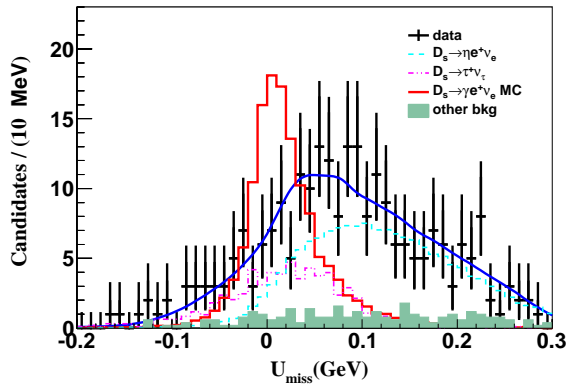}
  \caption{\label{fig:umiss} Distribution of $U_{\rm miss}$ for the selected $D_s^+\to\gamma e^+\nu_e$ candidates. The black points with error bars represent the data. The solid blue line corresponds to the overall fit, the magenta dashed-line histogram shows the background $D^+_s\to \tau^+ \nu_\tau$ and the cyan dashed-line histogram shows the background $D_s^+\to \eta e^+\nu_e$.  The solid red curve is from the simulated signal candidates normalized with a partial BF $\BR(D_s^+\to\gamma e^+\nu_e)= 7.5\times10^{-4}$.}
          \end{figure}

\begin{figure}
  \centering
  \includegraphics[width=8cm]{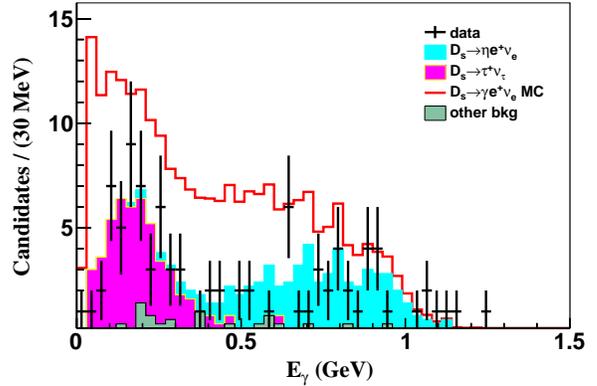}
  \caption{\label{fig:E_gamma}\small Energy spectrum of the radiative photon of selected candidates in the rest frame of an $e^+e^-$ system. The black points with error bars represent the data. The solid red curve shows the distribution of the simulated signal candidates normalized with a partial BF $\BR(D_s^+\to\gamma e^+\nu_e)$= $7.5\times10^{-4}$. An additional requirement of
$|U_{\rm miss}| < 0.06$ GeV has been imposed on the candidates shown in this plot.}
  \end{figure}

  Finally, the signal candidates are searched for in the data distribution of the kinematic variable
  \begin{equation}
  U_{\rm miss} \equiv E_{\rm  miss} - |\vec{p}_{\rm miss}|,
  \end{equation}
   where
   \begin{equation}
   E_{\rm miss} \equiv 2E_{\rm beam} - E_{\gamma} - E_e - E_{\rm ST} - E_{\gamma_{\rm soft}(\pi^0_{\rm soft})}
   \end{equation}
   and
   \begin{equation}
   \vec{p}_{\rm miss} \equiv -({\vec p}_\gamma+{\vec p}_e+{\vec p}_{\rm ST}+{\vec p}_{\gamma_{\rm soft}(\pi^0_{\rm soft})})
      \end{equation}
 in the $e^+e^-$ rest frame. Here, $E_i$ and $p_i$~$(i = \gamma_{\rm soft}(\pi^0_{\rm soft}), e^+$ or ST) are the energy and momentum of $\gamma_{\rm soft}(\pi^0_{\rm soft})$, positron and ST. The distribution of $U_{\rm miss}$ of the surviving DT candidates is shown in Fig.~\ref{fig:umiss}. The signal candidates of $D_s^+\to\gamma e^+\nu_e$ should peak around zero in the $U_{\rm miss}$ distribution, as shown by the signal MC sample~(black dashed line).
Figure~\ref{fig:E_gamma} shows the $E_\gamma$ distribution in the $U_{\rm miss}$ signal region $(-0.06,\,0.06)$~GeV, where the data points overlap with the simulated distributions of the backgrounds coming from the $D_s^+\to\eta e^+\nu_e$ and $D_s^+\to\tau^+\nu_\tau$ decays. No excess of signal candidates is observed in the signal region.

          \section{Result}

\begin{figure}[h]
     \centering
     \centering
  \includegraphics[width=8cm]{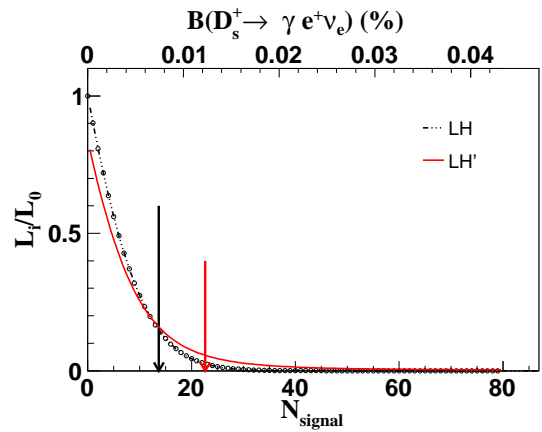}
  \caption{\label{fig:upper}\small Distribution of the normalized likelihood scan for $D^+_s\to\gamma e^+\nu_e$ candidates. The circles represent the maximum likelihood value when $\mathcal B(D^+_s\to \gamma e^+\nu_e)$ is fixed at the corresponding BF value. The black and red curves describe the smoothed likelihood curves before and after the inclusion of the systematic uncertainty. The black and red arrows show the corresponding upper limits of BF.}
    \end{figure}

     To measure the signal yield of the $D^+_s\to \gamma e^+\nu_e$ decay, an extended unbinned maximum-likelihood fit is performed to the $U_{\rm miss}$ distribution.  The result of the fit is shown as the solid line in Fig.~\ref{fig:umiss}. The signal shape is determined from the signal MC sample, and the numbers and shapes of the two backgrounds from the decays $\bkgEta$ with $\eta\to\gamma\gamma$ and $D^+_s\to\tau^+\nu_\tau$ with $\tau^+\to e^+\nu_e\bar \nu_\tau$
 are fixed by analyzing the corresponding MC sample. For the other background components, the shape is determined from the inclusive MC-simulated sample.  The DT efficiencies of the individual ST modes are listed in Table~\ref{tab:tag_yields}. Since no significant signal is observed, an upper limit on the BF of the $\sigGamma$ decay at the 90\% C.L. is set by solving the equation~\cite{Patrignani:2016xqp}
 \begin{equation}
 	\int_{0}^{{\mathcal B}^{\rm UL}} L(\BR)d\BR = 90\%.
 \end{equation}
A series of fits on the $U_{\rm miss}$ distribution is carried out, fixing the BF at different values. The resulting likelihood distribution $L$ is shown in Fig.~\ref{fig:upper}. The upper limit on the BF at the 90\% C.L. is found to be $\resultb$.

 The sources of systematic uncertainties that affect the upper limit calculation are discussed below.
 With the DT method, the systematic uncertainties related to the selection of the ST candidates are found to be negligible. To estimate the uncertainty in the ST yield and to avoid statistical fluctuations,
 a total of 1000 fits to generated samples have been performed by using alternative signal (double Gaussian function) and background (Chebyshev polynomial) shapes. The systematic uncertainties of 0.3\% and 0.2\% are obtained by taking the mean value of the distribution of the relative normalized difference between the pseudoexperiments and baseline fit results. The total systematic uncertainty in the ST tag yield is taken as the squared sum, and it is found to be 0.4\%.
 To estimate the systematic uncertainty due to not well-known-radiative photon due to the $D_s^+\to\gamma e^+\nu_e$ form factors, an alternative signal MC sample based on the single-pole model~\cite{Yang:2014rna} has been produced, the difference between the DT efficiency obtained with this model and the one with our nominal model at 0.025~GeV is 2.6\%, and the relative difference of fractions of the generated events in (0.01,\,0.025)~GeV between the two models is 8\%. Due to full correlation of the two systematic errors, they are added linearly to obtain the systematic uncertainty in the form factor model, 11\%.  The systematic uncertainties  attributed to the positron tracking and PID efficiencies are studied with a control sample of radiative Bhabha scattering events.
 The control sample and the $D_s^+\to\gamma e^+\nu_e$ simulation sample have different distributions in the momentum and angle of the positron.
 To account for these differences, a correction resulting from a two-dimensional reweighting in momentum and angle is applied to the positron tracking efficiency and to the positron PID efficiency. The total systematic error caused by uncertainties in positron tracking and PID is estimated to be 0.4\%.
The systematic uncertainty in the photon selection is evaluated using a control sample of $J/\psi\to\pi^+\pi^-\pi^0$ decays~\cite{Ablikim:2017zal}. It is determined to be 1.0\%. Systematic uncertainties of 1.1\% and 0.9\% due to the $E_{\gamma\,\rm extra}^{\rm max}$ and $N_{\rm char}^{\rm extra}$ selection criteria are estimated by analyzing the DT hadronic $D^{*+}_sD^-_s$ events. A systematic uncertainty of 0.3\% due to the FSR effect is computed by repeating the fit of the correction for the FSR effect, and taking the difference with respect to the baseline fit. 
The effect due to imperfect simulation of the $\chi^2_{\rm kine}$ distribution is estimated by repeating the likelihood scan via the $U_{\rm miss}$ fit with alternative $\chi^2_{\rm kine}$ requirements from 80 to 300 with a step of 5; the  largest difference of the BF upper limit to the baseline fit, 11\%,  is taken as a systematic uncertainty.

 To estimate the uncertainty of $U_{\rm miss}$ fitting related to the background shape, the fraction of each of the main background components is varied within one standard deviation of the corresponding BF~\cite{Patrignani:2016xqp}. The largest deviation with respect to the baseline result is 10\%. To avoid statistical fluctuations, a study based on pseudoexperiments is performed. A total of 1000 fits to generated samples is performed by varying the background shape. A systematic uncertainty of 10\% is obtained by taking the mean value of the distribution of the relative normalized difference between the pseudoexperiments and the baseline fit results.
  Differences between the tag of the ST modes in data and simulation are expected to impact the final result due to the different multiplicities. The associated systematic uncertainty is assigned as 0.5\% by studying the tracking/PID efficiencies and the photon selection in different multiplicities resulting in a difference between data and the MC sample.

Table~\ref{tab:sys_br} summarizes all the systematic uncertainties. The impact of the systematic uncertainty on the upper limit of the BF is taken into account by convolving the distribution of the sensitivity ($S$)
\begin{equation}
	LH^{'}(\BR) = \int_0^1LH\left(\frac{S}{\hat{S}}\BR\right)\exp\left(\frac{-(S-\hat{S})^2}{2\delta_S^2}\right)dS,
\end{equation}
where $LH(t) = C{\rm exp}\left(\frac{-(t-\hat{t})^2}{2\sigma_t^2}\right)$,  $C$ is a normalization constant, and $\hat{t}$ and $\sigma_t$ can be obtained when the likelihood distribution is fitted by $LH(t)$.   The value $\hat{S}$ is the nominal efficiency and $\delta_S$ is the systematic uncertainty on the BF~\cite{Stenson:2006gwf}. Finally, the upper limit on the BF of the $D_s^+\to\gamma e^+\nu_e$ decay is set to be $\result$ at the 90\% C.L..
\begin{table}
    \centering
    \caption{Systematic uncertainties in the determination of $\mathcal B(D_s^+\to\gamma e^+\nu_e)$.\label{tab:sys_br}}
    \begin{tabular}[t]{lc}\hline\hline
        Source  & Relative uncertainty~(\%)\\\hline
                    ST yields &0.4\\
    Form factor model& 11\\
    $e^+$ tracking \& PID &0.4\\
    Photon selection &1\\
    $E_{\gamma\,\rm extra}^{\rm max}$& 1.1\\
    $N^{\rm extra}_{\rm char}$& 0.9\\
    $\chi^2_{\rm kine}$ & 11\\
    FSR & 0.3\\
    $U_{\rm miss}$ fit & 10\\
    Tag bias& 0.5\\
    \hline
    Total&18.6 \\\hline\hline
    \end{tabular}
    \label{table:sys_br}
\end{table}

\section{Summary}

In summary, the first search for the radiative leptonic decay $D_s^+\to\gamma e^+\nu_e$ is performed using $e^+e^-$ collision data corresponding to an integrated luminosity of 3.19 fb$^{-1}$ collected at $\sqrt{s}=$ 4.178~GeV, by employing a DT technique. No significant signal for the signal decay $\sigGamma$ is observed.  With a 0.01~GeV cutoff on the radiative photon energy, the upper limit on the BF of the $D_s^+\to\gamma e^+\nu_e$ decay mode is set to be $\BR(\sigGamma)<$ $\result$ at the 90\% C.L.. The result is compatible with the theoretical predictions in Refs.~\cite{Lu:2002mn,Geng:2000if,Atwood:1994za,Burdman:1994ip}, but smaller than that in Ref.~\cite{Yang:2012jp}  which stated that the BF could be significantly enhanced by long-distance contribution.

\begin{acknowledgments}
The BESIII Collaboration thanks the staff of BEPCII and the IHEP computing center for their strong support. This work is supported in part by National Key Basic Research Program of China under Contract No.~2015CB856700;
National Natural Science Foundation of China (NSFC) under Contracts Nos. 11235011, 11335008, 11425524, 11625523, 11635010, 11875054, 11775027; the Chinese Academy of Sciences (CAS) Large-Scale Scientific Facility Program; the CAS Center for
Excellence in Particle Physics (CCEPP); Joint Large-Scale Scientific Facility Funds of the NSFC and CAS under Contracts Nos. U1332201, U1532257, U1532258; CAS Key Research Program of Frontier Sciences under Contracts Nos.
QYZDJ-SSW-SLH003, QYZDJ-SSW-SLH040; 100 Talents Program of CAS; National 1000 Talents Program of China; INPAC and Shanghai Key Laboratory for Particle Physics and Cosmology; German Research Foundation DFG under Contracts
Nos. Collaborative Research Center CRC 1044, FOR 2359; Istituto Nazionale di Fisica Nucleare, Italy; Koninklijke Nederlandse Akademie van Wetenschappen (KNAW) under Contract No.~530-4CDP03; Ministry of Development of
Turkey under Contract No.~DPT2006K-120470; National Natural Science Foundation of China (NSFC) under Contracts Nos. 11505034, 11575077; National Science and Technology fund; The Swedish Research Council;
The Knut and Alice Wallenberg Foundation (Sweden);
U. S. Department
of Energy under Contracts Nos.~DE-FG02-05ER41374, DE-SC-0010118, DE-SC-0010504, DE-SC-0012069; University of Groningen (RuG) and the Helmholtzzentrum fuer Schwerionenforschung GmbH (GSI), Darmstadt; WCU Program of National
Research Foundation of Korea under Contract No. R32-2008-000-10155-0; the Royal Society (United Kingdom).
\end{acknowledgments}


\end{document}